\documentclass[useAMS,usenatbib]{mn2e}

 \usepackage{graphicx}
 
\newcommand{\evh}[1]{{#1}}

\title[The distance to the Fornax dwarf spheroidal galaxy]{The distance to the Fornax dwarf spheroidal galaxy\thanks{Based on data collected at the European Southern Observatory,
         La Silla, Chile, Proposal No. 66.B-0615}\\}
\author[Rizzi et al.]{L.~Rizzi$^{1,2}$, E.V.~Held$^3$,I.~Saviane$^4$,R.B.~Tully$^1$,M.~Gullieuszik$^{3,5}$\\
$^1$Institute for Astronomy, University of Hawaii, 2680 
  Woodlawn Drive, HI 96822, USA \\
$^2$Joint Astronomy Centre, 660 N. A'ohoku Place, University Park, Hilo, HI 96720, USA\\
$^3$Osservatorio Astronomico di Padova, INAF, Vicolo Osservatorio 5, 
  I-35122 Padova, Italy\\
$^4$European Southern Observatory, 3107 Alonso de Cordova, Vitacura,
  Casilla 19001, Santiago 19, Chile\\
$^5$Dipartimento di Astronomia, Universit\`a di
  Padova, vicolo dell'Osservatorio 2, I-35122 Padova, Italy
         }

\begin{document}
 


\date{Received \dots; accepted \dots}

\maketitle
  
  \begin{abstract}
  A large multicolour, wide-field photometric database of the Fornax
  dwarf spheroidal galaxy has been analysed using three different
  methods to provide revised distance
  estimates based on stellar populations in different age intervals.
  The distance to Fornax was obtained from 
  the Tip of the Red Giant Branch measured by a new method, 
  and using the luminosity of Horizontal Branch stars and Red Clump stars corrected for stellar population effects.
  Assuming a reddening $E(B-V)=0.02$, the following distance moduli were
  derived: $(m-M)_0=20.71 \pm 0.07$ based on the Tip of the Red Giant
  Branch,
  $(m-M)_0=20.72 \pm 0.06$ from the level of the Horizontal Branch,
  and $(m-M)_0=20.73 \pm 0.09$ using the Red Clump method. The weighted 
mean distance modulus to 
  Fornax is $(m-M)_0=20.72 \pm 0.04$. All these
  measurements agree within the errors, and are fully consistent with
  previous determinations and with the distance measurements obtained in a
  companion paper from near-infrared colour-magnitude diagrams.
\end{abstract}


  \section{Introduction}

Along with the disrupted Sagittarius dwarf, Fornax is one of the most
massive satellites of the Milky Way. 
Fornax shows evidence of a complicated and extended star formation history and several  studies have contributed to clarify the variety of 
stellar populations that are present. This galaxy was one of the first to provide convincing evidence of the presence of a conspicuous amount of intermediate-age stars, probed both by the presence of luminous carbon stars \citep{1985ApJ...290..191A,1980ApJ...240..804A,1999Ap&SS.265..291A} and by the well populated red clump \citep{stet+1998,2000AA...355...56S}. Blue luminous stars populating the upper main sequence indicate that Fornax has been forming stars up to very recent times \citep{1999AJ....118.1671B,2004AJ....127..840P,2000AA...355...56S,stet+1998}. Finally, the detection of a significant number of RR Lyrae stars probes the presence of an old, metal-poor component \citep{2002AJ....123..840B,2005astro.ph..7244G}.
Most recently, stellar population gradients and stellar metallicities 
have been studied over a wide area of Fornax by \cite{batt+06}, confirming the presence of at least three distinct stellar components: a young population concentrated near the centre, an intermediate-age population, and an ancient population. The three components are found to be distinct from each other kinematically, chemically, and in their spatial distribution.

The distance to Fornax marks a reference point in the distance
ladder based on secondary distance indicators. Together with the Carina dwarf spheroidal and the Large and Small Magellanic Clouds, Fornax has been used as a benchmark to investigate possible discrepancies between different distance indicators, and to study their dependence on the properties of the underlying stellar populations \cite[e.g., see][]{2000ApJ...543L..23B,2003AJ....125.2494P}. Fornax is particularly suitable for these kind of investigations because of the contemporary presence of both old and intermediate-age/young stellar populations.

\evh{The distance to Fornax has} been determined using a number of
 different indicators, and the estimates range from $(m-M)_0=20.59 \pm
 0.22$ \citep{1985AA...152...65B} to $(m-M)_0=20.86$ \citep{2002AJ....123..789P}.
%
In particular, 
\cite{2000AA...355...56S} found $(m-M)_0=20.70 \pm 0.12$ based
 on the Tip of the Red Giant Branch (TRGB) technique, and $(m-M)_0=20.76
 \pm 0.04$ based on the luminosity of \evh{red} Horizontal Branch (HB)
 stars. \cite{2000ApJ...543L..23B} derived a slightly shorter distance
 modulus, $(m-M)_0=20.66$, based on TRGB
 and \evh{Red Clump (RC)} methods. 
%
\evh{A larger distance modulus, $(m-M)_0=20.86$, was obtained by 
\cite{2003AJ....125.2494P} using} $K$-band
 imaging of RC stars. A recent analysis of a wide-area database of near-infrared 
data by \cite{M.:fk} resulted in $(m-M)_0=20.74 \pm 0.11$
 from the mean magnitude of the Red Clump (RC), and $(m-M)_0=20.75 \pm
 0.19$ based on a population-corrected measurement of the TRGB in the
 $K$ band. \evh{Previous measurements of the distance to Fornax 
are summarised in Table~\ref{tab:literature}.}

\evh{The consistent results from the near-infrared photometry of
\cite{M.:fk} and the relatively large scatter of previous optical
distance estimates motivated us to revise the distance to Fornax 
by exploiting a large multi-band $BVI$ database of 
photometric observations of Fornax.  Our data, covering 1/4 square
degree and containing $\sim$ 450.000 stars, were obtained with the Wide
Field Imager at the ESO/MPG 2.2-m telescope at La Silla, Chile.
The range of distances found in the literature are based on different
data sets, each with its own filter set and photometric zero-point. 
Therefore, measuring the distance with different methods on a
single, homogeneous data set with high statistics may provide a standard
case of the {\it intrinsic uncertainties} in distance determinations in
stellar systems with extended star formation histories.

}

 The paper is organised as follows: Sect.~\ref{observations} presents the
 observations, the reduction techniques, and the resulting
 colour-magnitude diagram (CMD). The different methods used to derive
 the distance to the Fornax dwarf spheroidal are then presented and the
 results discussed in Sect.~\ref{trgb}, \ref{hb}, and \ref{rc}. Final
 remarks and discussion of the results are presented in 
 Sect.~\ref{discussion}.

\begin{table}
\caption{Fornax distance determinations}          
\label{tab:literature}      
\centering                          
\begin{tabular}{l l c l}        
\hline\hline                 
~~~Value & ~~Method & $E(B-V)$ & Reference\\    
\hline                        
$20.59 \pm 0.22$ & HB & 0.03   &  \cite{1985AA...152...65B}\\
$20.76 \pm 0.20$ & HB & 0.05  & \cite{1999AJ....118.1671B}\\
$20.76 \pm ...$ & HB & 0.05 & \cite{1990PASP..102..632D}\\
$20.70 \pm 0.12$ & TRGB & 0.03 & \cite{2000AA...355...56S}\\
$20.76 \pm 0.04$ & HB & 0.03 & \cite{2000AA...355...56S}\\
$20.65 \pm 0.11$ & TRGB & 0.03 & \cite{2000ApJ...543L..23B}\\
$20.66 \pm ...$ & RC & 0.03 & \cite{2000ApJ...543L..23B}\\
$20.86 \pm 0.013$ & RC & 0.03 & \cite{2003AJ....125.2494P} \\
$20.72 \pm 0.10$ & RR Lyrae & 0.04 & \cite{2005astro.ph..7244G} \\
$20.74 \pm 0.11$ & RC & 0.03 & \cite{M.:fk}\\
$20.75 \pm 0.19$ & TRGB & 0.03 &\cite{M.:fk}\\
\hline                                   
\end{tabular}
\end{table}
%

\section{Observations and Data Reduction}
\label{observations}
 
Wide field observations of Fornax 
were obtained with the WFI camera mounted at the
Cassegrain focus of the ESO/MPG 2.2-m telescope at La Silla, Chile. The
camera consists of eight 2k $\times$ 8k CCDs, closely mounted on a mosaic
pattern yielding a total field of view of 
$\sim 32\arcmin \times 32\arcmin$. 
The observations were carried out on October 21, 2001, as a
part of ESO proposal 66.B-0615. Both long and short exposures were
obtained: $3 \times 900 + 1 \times 120 $ seconds in $B$, $2 \times 900 +
1 \times 120$ seconds in $V$, and $2 \times 900 + 1 \times 120$ seconds
in $I$.  The
absolute calibration was obtained by observing stars from the
list of \cite{1992AJ....104..340L} \evh{on each CCD}.
Pre-reduction was performed within the
IRAF\footnote{The Image Reduction and Analysis Facility (IRAF) software
is provided by the National Optical Astronomy Observatories (NOAO),
which is operated by the Association of Universities for Research in
Astronomy (AURA), Inc., under contract to the National Science
Foundation.} environment, \evh{using} the {\sc MSCRED} package by
\cite{1998ASPC..145...53V}. The subsequent reduction steps 
\evh{made use of the {\sc WFPRED} script package} 
developed by two of us (L.R. and E.V.H.) at
the Padova Observatory. This package effectively deals with the problems
of astrometric and photometric calibration. Images were astrometrically
calibrated using a polynomial solution computed on star fields taken
from the list of \cite{1999AJ....118.2488S}. 
%
%
Crowded field stellar photometry was performed with the {\sc daophot ii
/ allstar} package \citep{1987PASP...99..191S}, 
\evh{with point-spread
functions (PSFs) independently computed} for each CCD and each
filter. \evh{Aperture corrections were used to match the PSF
photometry with the aperture photometry, using growth-curve analysis of
bright isolated stars.  Finally, the photometric catalogues were
independently calibrated for each CCD using the observations of standard
stars.  The RMS of the photometric calibration is $\sim 0.02$ mag.}
An extensive set of artificial star experiments
\evh{was} run to estimate the degree of completeness and the
distribution of photometric errors. Completeness levels drop to 10\% at
$V\sim24.5$ where the photometric errors \evh{reach} 0.25 mag.

\begin{figure}
\includegraphics[scale=0.4]{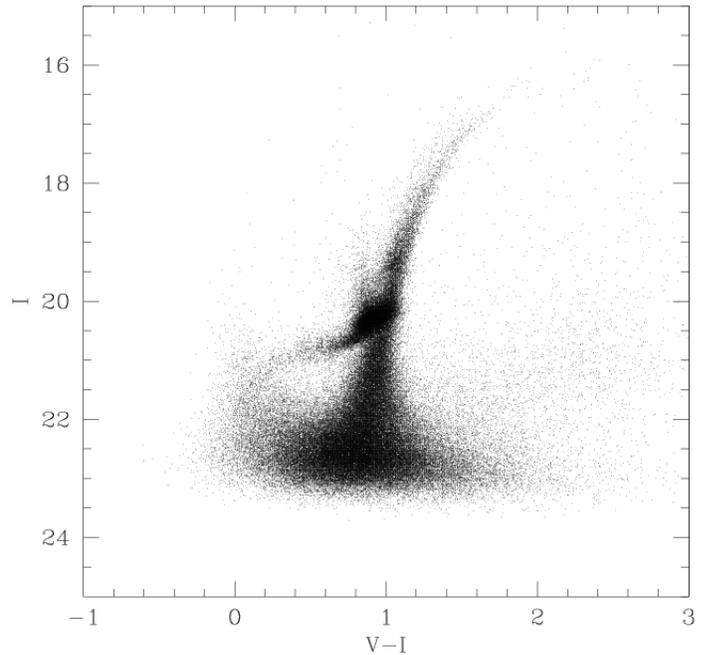}
\caption{$(V-I,I)$ colour magnitude diagram of Fornax. 
Only stars with photometric error less than 0.2 \evh{mag} are shown.}
\label{cmds1}
\end{figure}


\evh{The resulting $(V-I)$, $I$ CMD is presented in Figs.~\ref{cmds1}. 
The limiting magnitude is around $V \sim 24$, corresponding} to the
position of the old main sequence turn-off. Note that the figure only
shows stars with a photometric error less than 0.2, and is
consequently limited to $V \approx$ 23.5. A number of \evh{known
features are evident: the} well populated luminous main sequence of
stars as young as 0.3 Gyr, the wide and non-uniformly populated red
giant branch (RGB), the conspicuous red clump of intermediate-age
stars in their helium burning phase, and the less populated but
evident horizontal branch \evh{\cite[see,
e.g.,][]{stet+1998,2000AA...355...56S, 2004AJ....127..840P,batt+06}}.

\evh{These
features are used} in Sect.~\ref{rc} to derive information on
the star formation history. \evh{This CMD 
offers us the possibility to use several different methods to
estimate the distance to Fornax}. The TRGB is bright, well defined and well
populated, with 
\evh{manageable AGB contamination}. The RC is
prominent \evh{and bright} enough to be usefully adopted as a distance
indicator. The HB is not quite as populated as the RC, \evh{but still}
contains enough stars to derive a precise average magnitude.

\section{Distance based on the Tip of the Red Giant Branch}
\label{trgb}

%

\evh{In low mass stars, the He ignition occurs under degenerate
conditions at almost the same luminosity, with very little dependence on
metallicity or age. The observational evidence of this physics}
is a sharp cut-off of the luminosity function of the RGB,
approximatively located at $M_I \sim -4$. \cite{1990AJ....100..162D} and
\cite{1993ApJ...417..553L} demonstrated the power of this distance indicator applied
to nearby galaxies. \cite{1993ApJ...417..553L} provided both an absolute
calibration of the $I$-band luminosity of the TRGB ($M_I^{\rm TRGB}$) and an
objective method to estimate its position on a CMD, based on a digital
Sobel filter.
\evh{This technique was refined by \cite{1996ApJ...461..713S},who replaced the binned} luminosity function with an adaptively smoothed
probability distribution.  \cite{2002AJ....124..213M} introduced a new
way to estimate the position of the TRGB, based on a maximum-likelihood
approach that effectively \evh{uses} 
all the stars around the tip region. In a
series of papers, \cite{1999AJ....118.1738F, 2000AJ....119.1282F} and
 \cite{2001ApJ...556..635B, 2004AA...424..199B} obtained a new
robust calibration of the magnitude of the tip, extended to \evh{higher}
metallicities (up to $\rm{[Fe/H]} = -0.2$) and to infra-red pass-bands.

\begin{figure}
\centering
\resizebox{\hsize}{!}{\includegraphics[angle=270]{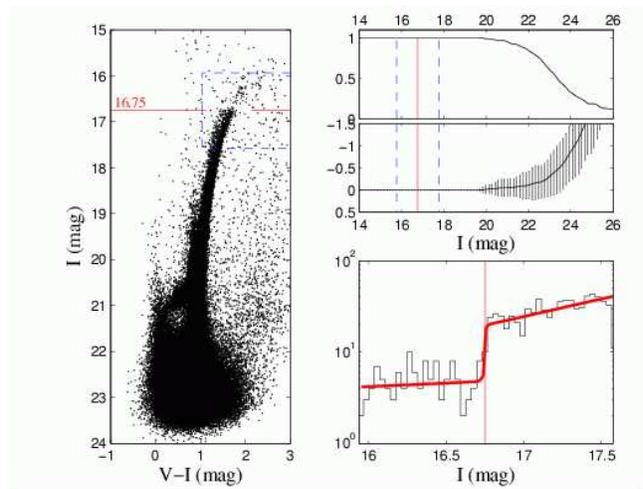}}
\caption{\evh{ Detection of the RGB tip using the maximum-likelihood
method of \citet{2006astro.ph..3073M}. The TRGB is indicated in the left
panel, while the right panels show the completeness function (top
panel), the distribution of photometric errors (middle panel), and the
luminosity function near the RGB tip (bottom panel). The observed LF is
indicated by the {\it solid (black)} histogram, while the {\it thick
(red)} line shows the best fitting model.  }}
\label{fig:trgb}
\end{figure}

\evh{A modified and optimised version of the maximum-likelihood approach
has been recently presented by \citet{2006astro.ph..3073M}. The new method
has the advantage that} completeness, photometric errors, and biased
error distributions are fully taken into account. 
The application of this method to the CMD of Fornax is presented in
Fig.~\ref{fig:trgb}. The left panel of Fig.~\ref{fig:trgb} shows the
CMD and the limits of the \evh{CMD region} selected for the fitting
procedure.  
\evh{ On the right half of the figure the completeness function and the
distribution of photometric errors derived from artificial star
experiments are shown.  A biased error distribution is evident at
magnitudes fainter than $I=24$. The lower panel shows the result of the
fit.}

The tip is detected at $I^{\rm TRGB}=16.75 \pm 0.02$, and the average colour
of stars at the tip is $(V-I) = 1.64 \pm 0.03$. As a first approximation, the colour of stars at the tip can be used to estimate the metallicity of the underlying stellar population. By using the relation
presented in \citet{2001ApJ...556..635B} (their Figure~1), 
\evh{this colour is 
converted} into a metallicity $\rm{[Fe/H]} = -1.50 \pm 0.04$ (the
error is purely statistical and does not contain any systematic
contribution). Using this value \evh{and the TRGB calibration of
\cite{2001ApJ...556..635B} (their equation 4)}, we obtain $M_I^{\rm TRGB} =
-4.07$.
\evh{The corrected distance is then derived by adopting 
the reddening value $E(B-V)=0.02$ from the 
infra-red dust maps of \cite{1998ApJ...500..525S} and the
relation $A_I = 1.94 E(B-V)$. 
Using the relation
$$ (m-M)_0=I^{\rm TRGB}-M_I^{\rm TRGB}-1.94\times E(B-V)$$
the distance to Fornax from the RGB tip is then} $(m-M)_0=20.78 \pm 0.04$.

The main contribution to the error affecting this determination comes from the conversion
of the mean colour at the level of the tip to metallicity. In the case of Fornax, this 
conversion is not actually necessary. Indeed, several determinations of the metallicity of Fornax exist in \evh{the} 
literature. \cite{2000AA...355...56S} found that the average metallicity
is $\rm{[Fe/H]} \sim -1.0 \pm 0.15$ 
\citep[on the scale of ][]{1984ApJS...55...45Z},
with tails extending to $-2.0 <
\rm{[Fe/H]} < -0.7$.  They also concluded that a model involving two
populations seems to provide a good description of the star content of
this galaxy, with the older populations having $\rm{[Fe/H]}=-1.82$. With a similar technique, 
applied to $V-K$ colours \cite{M.:fk} found a mean age-corrected mean metallicity
of $[M/H] \simeq -0.9$.
\cite{2001MNRAS.327..918T} found $\langle\rm{[Fe/H]}\rangle = -1.0$, later
confirmed with high resolution spectroscopy
\citep{2003AJ....125..707T}. \evh{More recently, 
\cite{2004AJ....127..840P} and \cite{batt+06} 
derived the metallicity distribution of Fornax RGB stars from
spectroscopy in the CaII triplet region. Both studies agree on finding
a metallicity distribution centred at $\rm{[Fe/H]} \simeq -0.9$
\citep[on the scale of][]{1997AAS..121...95C} with tails extending to
$\rm{[Fe/H]} \sim -2.2$ and $\rm{[Fe/H]} \sim -0.2$.} In summary, both
photometric and spectroscopic determination agree in suggesting a
metallicity $\rm{[Fe/H]} \sim -1.0$. Using this value, rather than the
value derived from direct conversion of the average colour at the tip
into metallicity, we derive $M_I^{\rm TRGB}=-4.00$, and the best
estimate of the distance based on the TRGB method is then
$(m-M)_0=20.71 \pm 0.07$. The error we quote is quite large, and is
derived by estimating the difference between this last distance
measurement and the value derived by simple conversion between colour
at the tip and metallicity, considering that the metallicity
uncertainty is the dominant factor in the error estimate.

\section{Distance based on Horizontal Branch stars}
\label{hb}

\begin{figure}
\centering
\resizebox{\hsize}{!}{\includegraphics{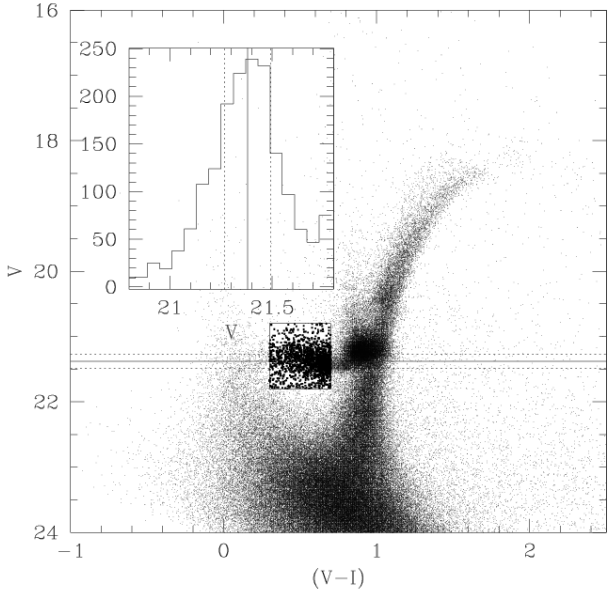}}
\caption{\evh{ 
Measurement of the level of the horizontal branch in Fornax. The
magnitude histogram of the stars in the outlined CMD region (most of
which are RR Lyrae variable stars) is shown in the inset. Solid and
dotted lines indicate the mean $V$ magnitude of HB stars 
and the r.m.s. of the data, respectively. 
}}
\label{fig:hb}
\end{figure}

\evh{The average luminosity of RR Lyrae and HB stars is the most widely
used Pop.~II distance indicator.  Our measurement of the HB level in
Fornax} is presented in Fig.~\ref{fig:hb}. Stars with $0.3 < V-I < 0.7$
and $20.8 < V < 21.8$ were selected, and their luminosity function
computed. The average \evh{magnitude of these stars is $V_{\rm HB}= 21.38
\pm 0.04$, where the error was inferred from the typical magnitude error
at the corresponding level. Errors caused by the binning in magnitude,
estimated by adding different shifts to our choice of bins, were found
to be negligible ($<0.001$ mag).}

\evh{Assuming an absolute magnitude for HB stars (actually, 
RR Lyrae stars) requires knowledge of metallicity. 
A lot of work has been devoted to determine the metallicity
dependence of the luminosity of RR\,Lyrae stars.  In this paper, we
adopt the recent calibration of} \citet{2003LNP...635..105C}:
$$ M_V(HB) = (0.23 \pm 0.04)(\rm{[Fe/H]+1.5)} + (0.62 \pm 0.03)$$
We also consider the previous 
calibration from \cite{2000ApJ...533..215C}.
%
In both cases, the metallicity is on the scale of
\citet{1997AAS..121...95C}.

%
In the previous Section we mentioned a number of estimates of the metallicity of Fornax, based
on the RGB stars.
\evh{In this Section, however,  the metallicity we are interested in is that of
the old stellar population producing the HB and RR Lyrae stars.}  

A search for variable stars by \cite{2002AJ....123..840B} resulted in an
average metallicity of the variables $\rm{[Fe/H]}=-1.64$ \evh{on} the
Butler-Blanco scale, which is known to be more metal-rich than the scale
of \cite{1984ApJS...55...45Z}.  This value is confirmed by
preliminary results of a study of RR Lyrae in the field of Fornax by
\cite{2005astro.ph..7244G}, suggesting a metallicity $\rm{[Fe/H]}=-1.78$ on the scale of \cite{1997AAS..121...95C}.
\evh{
This range of results clearly reflects the complex enrichment
history of Fornax. For the absolute calibration of HB stars,  
we assume that HB stars belong to the oldest and most
metal-poor stellar population in the galaxy. For this reason, and
taking into account the results of 
\cite{2000AA...355...56S},  
\cite{2002AJ....123..840B}, and 
\cite{2005astro.ph..7244G}, we adopted} a metallicity of HB
stars $\rm{[Fe/H]} \simeq -1.8$ \evh{on} the scale of
\cite{1984ApJS...55...45Z}, corresponding to $\rm{[Fe/H]} \simeq -1.6$ on the
scale of \cite{1997AAS..121...95C}.

The resulting distance modulus \evh{was then derived as:}
$$(m-M)_0=V_{\rm HB}-M_V^{\rm HB}-3.315 \times E(B-V)$$

Using the calibration of \citet{2003LNP...635..105C} the absolute
magnitude of HB stars is $M_V^{\rm HB}=0.59 \pm 0.04$ and the 
\evh{absorption-corrected distance}  is
$(m-M)_0=20.72 \pm 0.06$.
\evh{The calibration of \cite{2000ApJ...533..215C} would yield 
$M_V^{\rm HB}=0.52 \pm 0.04$ and a resulting distance modulus $(m-M)_0=20.79
\pm 0.04$.}

It is well known that RR Lyrae stars spend most of the time near the
faint limit of their light curve, so that a straight average of the
magnitudes measured in a single epoch might be biased towards fainter
values (resulting in an overestimate of the distance). To quantify
this possible source of error, we can look at the average magnitude of
RR Lyrae stars as measured by \citet{2002AJ....123..840B} and
\citet{2005astro.ph..7244G}. In both cases, the
reddening-corrected magnitude is $\langle V_0\rangle\sim 21.27$, or $\langle V\rangle$=21.34 without
reddening correction. This is 0.05 mags brighter than the values we computed, and this is 
probably a reliable estimate of the errors related to averaging single-epoch 
observations rather than the complete light curve.

\section{Distance based on the Red Clump method}
\label{rc}
 

Core helium-burning stars of intermediate age form a well defined clump
of stars at a magnitude slightly brighter than the HB and near the
Hayashi line. This clump has been claimed to provide a very
accurate standard candle \citep[e.g., see][ and references
therein]{2000ApJ...531L..25U}.  Once the mean $I$-band magnitude of the
red clump, $I^{\rm RC}$, is known, the absolute distance modulus can be
derived using the relation:

$$ (m-M)_0=I^{\rm RC}-M_I^{\rm RC}-A_I-\Delta M_I^{\rm RC} $$

where $M_I^{\rm RC}$ is the 
mean absolute magnitude measured for nearby red clump stars 
whose distances were measured using independent techniques (such as
trigonometric parallaxes for the Hipparcos sample), $A_I$ is the
interstellar absorption in the direction of the object and $\Delta
M_I^{\rm RC}$ is a population correction term 
accounting for the different mixture of stellar ages in 
the local sample of stars and the Fornax galaxy.
The correction term
$\Delta M_I^{\rm RC}$ was initially neglected
\citep{1998ApJ...494L.219P,1998ApJ...500L.141S,1998AcA....48....1U}, but
\cite{1998ApJ...500L.137C} and \cite{1998MNRAS.301..149G} pointed out
that it is non-negligible according to theoretical models of clump
stars. 
Empirical determinations of the dependence of $M_I^{\rm RC}$ from
stellar parameters by \cite{1998AcA....48..383U,1998AcA....48..113U}
give a linear relation between $M_I^{\rm RC}$ and $\rm{[Fe/H]}$ with little
or no dependence from stellar ages. 
\cite{2000AcA....50..279U} suggested the
following relation for the RC luminosity:
$$ M_I^{\rm RC}=(0.14 \pm 0.04) \times ({\rm [Fe/H]}+0.5)-0.29 \pm 0.05 $$
Discussing the problem from a theoretical point of view,
\cite{2001MNRAS.323..109G} found a clear and non linear dependence of
$M_I^{\rm RC}$ both from age and metallicity.  They determined the mean
value of the red clump stars by averaging the contribution from all the
stars in the core helium burning phase, \evh{and} provided tables of
$M_I^{\rm RC}$ as a function of age for 6 different metallicities ranging
from $Z=0.0004$ to $Z=0.03$.
Following their precepts, we computed the expected correction to
the absolute magnitude of the red clump in Fornax, given its
star-formation history  and chemical enrichment law.


The apparent magnitude of the RC in Fornax was measured by 
selecting stars with $ 0.8 < (V-I)_0 < 1.1 $ and $ 19.5 <
I_0 < 21.0$.  After this selection the luminosity function was computed,
a linear continuum subtracted, and Gaussian fitting used to measure the
average magnitude of RC stars.
%
\begin{figure}
\centering
\resizebox{\hsize}{!}{\includegraphics{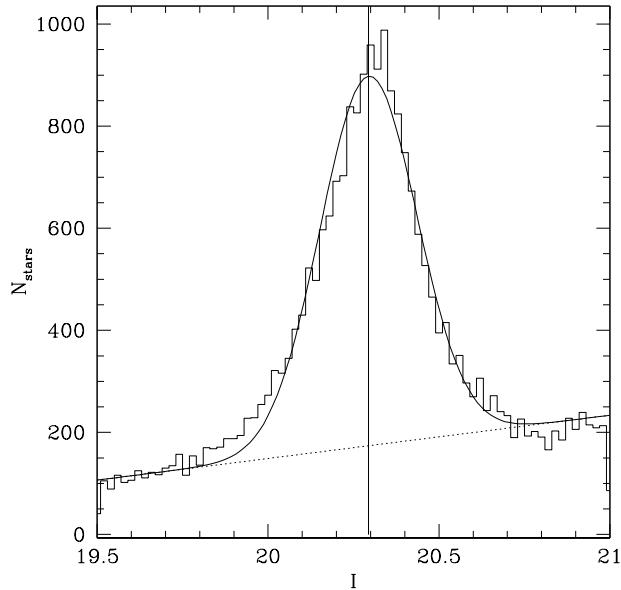}}
\caption{
Measurement of the mean $I$ magnitude of red clump stars 
in Fornax. 
}
\label{fig:rc}
\end{figure}
%
According to our measurements, illustrated in Fig.  \ref{fig:rc},
the Fornax RC is found at $I^{\rm RC}=20.29 \pm 0.03$.
The errors are inferred from the typical magnitude error at the
corresponding level. Errors due to the magnitude binning were estimated
by adding a sequence of shifts to the zero-point of the binning.  Such
errors were found to be less than $0.001$ and were consequently
neglected.

We now need to determine the population correction to correct for the difference between the absolute luminosity of the RC in the Hipparcos sample and in Fornax. To derive the population correction we follow these steps: (1) we perform a full inversion of the CMD of Fornax, to derive its star formation history, (2) we measure the luminosity of RC stars on a simulated diagram that closely reproduces the star formation history of Fornax ($M^{\rm RC}_{I,Fornax}$), (3) we construct a simulated CMD based on literature studies of the age and metallicity distribution of RC stars of the Hipparcos sample ($M^{\rm RC}_{I,Hipparcos}$), (4) we compute the population correction $\Delta M_I^{\rm RC}=M^{\rm RC}_{I,Fornax}-M^{\rm RC}_{I,Hipparcos}$.

\begin{figure}
\centering
\resizebox{\hsize}{!}{\includegraphics{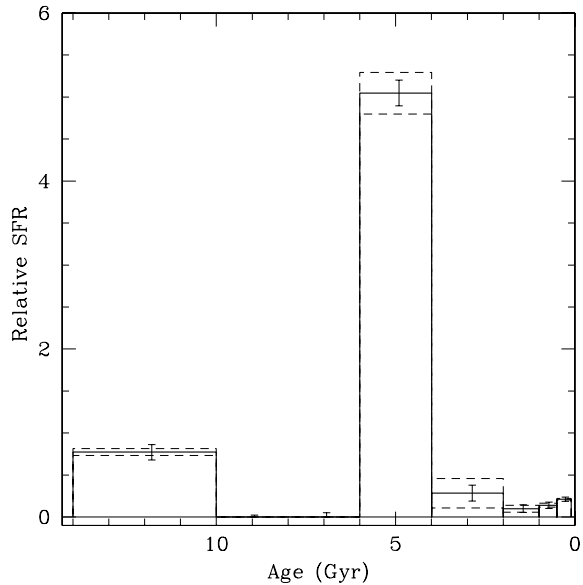}}
\caption{
The star formation history of Fornax dwarf spheroidal adopted 
in this paper. The star
formation rate is normalised to an average value of 317
$M_\odot$/Myr. Vertical error bars indicate the 1 $\sigma$ confidence
level returned by the best-fit algorithm, while dashed lines
show the 1 $\sigma$ uncertainty derived from an extensive set of 
Monte Carlo simulations.}
\label{sfh}
\end{figure}

%
To derive the star formation history of Fornax, we used the chemical enrichment law of \cite{2004AJ....127..840P} and
the CMD simulation technique presented in
\cite{2002ASPC..274..490R,2003ApJ...589L..85R} to perform a full
inversion of our Fornax CMD. The result is shown in Fig.~\ref{sfh}. Note that there is a
very good agreement between the SFH derived here and the one presented
in \cite{2003AJ....125..707T}.
Using the results of this simulation, we applied the above
described selection of RC stars to the simulated diagram, and derived
an absolute magnitude of the RC that fully takes into account the
chemical evolution history and the SFH of the galaxy.  Gaussian
fitting to the background-subtracted luminosity function of simulated
stars in the RC region results in $M_{I,Fornax}^{\rm RC}=-0.42 \pm 0.03$.
The error was derived by Monte Carlo simulations, by repeating 500
times the CMD inversion using a completely new set of simple stellar
populations. 

To derive the absolute magnitude of the RC in the Hipparcos sample, we used the star formation history and chemical enrichment laws presented in \cite{2000A&A...358..850R,rocha}. The same magnitude and colour selection was adopted as in the case of the simulated CMD of Fornax, and the fitting routine yielded the value: $M_{I,Hipparcos}^{\rm RC}=-0.17 \pm 0.06$. Note that this is same value derived by \cite{2001MNRAS.323..109G}. 

\cite{1998ApJ...500L.141S} estimate the mean absolute magnitude of RC stars in the Hipparcos sample: $M_I^{\rm RC}=-0.23 \pm 0.03$.
Using all these values
%
%
we find a distance $(m-M)_0=20.73 \pm 0.09$.

\evh{For comparison, we followed 
a different approach and used the relation derived by
\cite{2000AcA....50..279U} to estimate the distance to Fornax from 
$M_I^{\rm RC}$, by adopting an average 
metallicity  $\rm{[Fe/H]} = -1.0$. In this case, $M_I^{\rm RC}=-0.36$ and  
the distance modulus is $(m-M)_0=20.61 \pm 0.05$.}

\section{Summary and conclusions}
\label{discussion} 

%
\begin{table}
\caption{Distance to Fornax}          
\label{tab:summary}      
\centering                          
\begin{tabular}{lcc}        
\hline\hline                 
Method & Distance & Distance\\
              & E(B-V)=0.02 & E(B-V)=0.05\\
\hline                        
RGB tip   & $20.71 \pm 0.07$ & $20.65 \pm 0.07$\\
HB & $20.72 \pm 0.06$ & $20.62 \pm 0.06$\\
Red clump        & $20.73 \pm 0.09$ & $20.67 \pm 0.09$\\
\hline
Mean      & $20.72 \pm 0.07$ & $20.65 \pm 0.07$\\
Weighted Mean & $20.72 \pm 0.04$ & $20.64 \pm 0.04$\\
\hline                                   
\end{tabular}
\end{table}
%


\evh{In this paper we have presented a revision of the distance to
Fornax based on wide-field $BVI$ observations, and using } three different
distance indicators, namely the Tip of the Red Giant Branch, the
Horizontal Branch, and the Red Clump.  \evh{Our measurements are
summarised in Table~\ref{tab:summary} for two different values of the
interstellar reddening, $E(B-V)=0.02$ and $E(B-V)=0.05$, 
encompassing the range
of values adopted in previous works. Within the errors the different 
methods agree.}
The straight average of the measurements is $(m-M)_0=20.72
\pm 0.07$ for $E(B-V)=0.02$. 

It is worth discussing the individual results of each method
\evh{by comparing them  with previous determinations.
Previous measurements} based on the TRGB in the $I$ band were obtained
by \cite{2000AA...355...56S}, who found $(m-M)_0=20.70 \pm 0.04$, and
\cite{2000ApJ...543L..23B}, who found $(m-M)_0=20.65 \pm 0.11$. The
agreement with the measurement presented in this paper is fairly good,
but even better if we only look at the observable, which is the
magnitude of the TRGB. Indeed, \cite{2000AA...355...56S} found
$I^{\rm TRGB}=16.72$, and \cite{2000ApJ...543L..23B} found $I^{\rm TRGB}=16.73$,
almost coincident with the value derived here, $I^{\rm TRGB}=16.75$. The
difference in the derived distances are entirely due to different
assumptions \evh{on the absolute magnitude of the tip, the metallicity,
and the reddening}.

Previous determinations of the distance to Fornax based on the HB are
presented in \cite{2000AA...355...56S}, who found $(m-M)_0=20.76 \pm
0.04$, and in  \cite{2005astro.ph..7244G}, who found $(m-M)_0=20.72 \pm 0.10$. 
Both values are fully consistent with the values presented in this
paper, as is the agreement of the observable, the
mean level of the HB stars. \cite{2000AA...355...56S} found
$V_{\rm HB}=21.37 \pm 0.04$ while the value measured in this paper is
$V_{\rm HB}=21.38 \pm 0.11$.
\evh{Note that the \cite{2000AA...355...56S} value refers to
  the red part of the HB, while the present measurement refers to the
  CMD region populated by RR\,Lyrae variable stars. 
This value is in perfect agreement with the mean
  luminosity of RR\,Lyrae stars in the Fornax globular cluster \#4 
determined from a different and independently calibrated
data set by \cite{grec+06apj}.}

Finally, a previous determination of the distance to Fornax based on the
optical luminosity of RC stars is presented in 
\cite{2000ApJ...543L..23B}, who finds $(m-M)_0=20.66$ based on a
\evh{magnitude of the RC 
$I^{\rm RC}=20.31$,} almost coincident with the value derived here,
$I^{\rm RC}=20.29$.  The determination presented in \cite{2003AJ....125.2494P}, $I^{\rm RC}=20.33$, is also fully consistent with our measurement.
It is interesting to note that the difference $I_{\rm RC}-I^{\rm TRGB} = 3.54 \pm 0.04$ found in this paper 
is in good agreement with the linear fit presented in Figure 4 of \cite{2000ApJ...543L..23B}, obtained by adopting the RC calibration
of \cite{2000ApJ...531L..25U}.

\evh{We conclude that the different data sets recently published show 
an almost perfect agreement with each other and with the results obtained in this paper.
On the other side, there are still small differences between the distance moduli obtained with different methods and by different authors. The differences are mainly due
to the different details in the choice of the standard candle, while there is a 
striking agreement between the different photometry sets. The main
sources of uncertainty are (a) the relatively uncertain metallicity, (b)
the still debated absolute calibration of these methods, and (c) the
adopted reddening. It is on these open subtle calibration issues that
the attention should focus to bring the different distance measurements
in perfect agreement.}


\section*{Acknowledgements} 
We would like to thank the referee, D. Bersier, for his useful and constructive  
suggestions that significantly improved the paper.

 
\end{document}